# Cross-Layer Multi-Cloud Real-Time Application QoS Monitoring and Benchmarking As-a-Service Framework

Khalid Alhamazani, Rajiv Ranjan, Prem Prakash Jayaraman, Karan Mitra, Chang Liu, Fethi Rabhi, Dimitrios Georgakopoulos, Lizhe Wang

**Abstract**— Cloud computing provides on-demand access to affordable hardware (e.g., multi-core CPUs, GPUs, disks, and networking equipment) and software (e.g., databases, application servers and data processing frameworks) platforms with features such as elasticity, pay-per-use, low upfront investment and low time to market. This has led to the proliferation of business critical applications that leverage various cloud platforms. Such applications hosted on single/multiple cloud provider platforms have diverse characteristics requiring extensive monitoring and benchmarking mechanisms to ensure run-time Quality of Service (QoS) (e.g., latency and throughput). This paper proposes, develops and validates CLAMBS—Cross-Layer Multi-Cloud Application Monitoring and Benchmarking as-a-Service for efficient QoS monitoring and benchmarking of cloud applications hosted on multi-clouds environments. The major highlight of CLAMBS is its capability of monitoring and benchmarking individual application components such as databases and web servers, distributed across cloud layers (*-aaS), spread among multiple cloud providers. We validate CLAMBS using prototype implementation and extensive experimentation and show that CLAMBS efficiently monitors and benchmarks application components on multi-cloud platforms including Amazon EC2 and Microsoft Azure.

—————————— ◆ ——————————

## 1 INTRODUCTION

CLOUD computing has emerged as a successful computing paradigm and has revolutionized the way computing infrastructure is virtualized and used [1]. It offers a flexible access to huge pool of virtually infinite resources such as processing, storage and network with practically no capital investment and modest operating costs, proportional to the actual use (pay-as-you use model) [2]. The elasticity, pay-as-you-go model and low upfront investment offered by clouds, have led to the proliferation of number of application providers. For example, popular applications such as Netflix and Spotify use clouds such as Amazon EC2 to offer their services to the millions of customer's worldwide.

The success of cloud computing can be attributed to virtualization that enables multiple instances of virtual machines (VMs) to run on a single physical machine via resource (CPU, storage and network) sharing. Thereby, leading to flexibility and elasticity, as new instances can be launched and terminated as and when required. Further, virtualization also leads to higher security as multiple instances running on a VM runs independently of each other [20].

The cloud platform is logically composed of three layers. These include: Software-as-a-Service (SaaS), Platform-as-a-Service (PaaS) and Infrastructure-as-a-Service (IaaS). For example, applications such as email and games are hosted on SaaS layer; applications such as databases and web servers are hosted on the PaaS layer; and finally, IaaS include resources such as VMs, network and CPU resources. For the efficient use of cloud resources and to meet service level agreements (SLAs), it is imperative that applications and components deployed across all these layers (*aaS) and possibly distributed across multiple clouds are monitored at runtime and are benchmarked [26]. In particular, application developers, system designers, engineers and administrators have to be aware of the compute, storage, networking resources, application performance and their respective quality of service (QoS) across all the cloud layers; as QoS parameters including latency and throughput play a critical role in upholding the grade of services delivered to the end customers based on the agreed upon SLAs.

In a cloud computing system, the QoS parameter values are stochastic and can vary significantly based on unpredictable user workloads, hardware and software failures. Thereby, necessitating the awareness of system's current software and hardware service status such that QoS targets of cloud-hosted applications are met [21]. Cloud monitoring and benchmarking can assist in the holistic monitoring and awareness of applications and

————————————————

- *R. Ranjan and P.P. Jayaraman are with the CSIRO Digital Productivity, Building 108 North Road, Acton-2601, Australia. E-mail: {rajiv.ranjan, prem.jayaraman} @ csiro.au.*
- *C. Liu with University of Technology Sydney, Australia. E-mail: {chang.liu, Jinjun.chen}uts.edu.au*
- *K. Alhamazani and F. Rabhi are with the School of Computer Science and Engineering, University of New South Wales. E-mail: {ktal130,Fethir} @ cse.unsw.edu.au.*
- *K. Mitra is with Luleå University of Technology, Skellefteå Campus, 93187 Skellefteå, Sweden. E-mail:karan.mitra@ltu.se.*
- *D. Georgakopoulos is with Royal Melbourne Instituteo of Technology, Melbourne Australia. Email: dimitrios.georgakopoulos@rmit.edu.au*
- *L. Wang is with the Chinese Academy of Sciences, Beijing, China. E-mail: lizhe.wang@gmail.com.*

components at *aaS layers to meet SLAs [21][23][24][25]. Monitoring is required for [26]: (i) QoS management of software and hardware resources; (ii) runtime awareness of the applications and resources for cloud providers and application developers/administrators; and (iii) detecting and debugging software and hardware problems affecting applications' QoS. Additionally, benchmarking can be used for: (i) understanding application performance (resource and network) before application deployment; (ii) facilitating application base lining; and (iii) enabling continual comparison of applications QoS performance against baseline results. Recently, both industry and academia has focused on cloud monitoring and benchmarking [27-32]. However, most of the approaches are limited to one cloud provider and/or one cloud layer (IaaS/PaaS/SaaS).

We assert that in a distributed application hosting environments such as clouds, there is a need for application deployment across multi-cloud providers and multi-layered environments to benefit from resilience and economies of scale. This necessitates QoS monitoring and benchmarking at multiple cloud service layers. For example, the failure of a particular VM (IaaS layer) affects the QoS of web application (PaaS layer) or database application (PaaS layer) hosted within that VM. This ultimately affects the QoS of end-user of that web application offering (SaaS layer). This establishes the need for cloud monitoring and benchmarking framework that is capable of monitoring applications and components across multiple cloud layers and across multiple cloud provider environments [36]. Further, benchmarking aids in ensuring that the system's current performance is as good as its baseline performance.

A multi-layer and multi-cloud monitoring and benchmarking system can enable cloud providers and application developers to efficiently manage cloud resources and application components by gaining an in-depth understanding of the QoS parameter values across cloud layer in a multi-cloud setting. The current cloud-application monitoring frameworks such as Amazon CloudWatch[1] typically monitor the entire VM as a black box. This means that the actual behavior of each application's component is not monitored separately. This renders application monitoring with a limited scope where not all components distributed across PaaS and IaaS layers are monitored and benchmarked holistically. This limiting factor reduces the ability for fine-grained application monitoring and QoS control across layers. Further, current cloud monitoring frameworks are mostly incompatible across multiple cloud providers. For example, Amazon CloudWatch does not allow monitoring application components hosted on non-AWS platforms. This defeats the distributed nature of cloud application hosting. These drawbacks trigger the significance of having interoperable and multi-layer enabled monitoring techniques and frameworks. Finally, current approaches lack the ability to benchmark application performance deployed different layers allowing the service provider to establish baseline performance estimates.

**Contribution**: The key contribution of this paper is to address an important challenge of cross-layer cloud monitoring and benchmarking in multi-cloud environments. In particular, we propose, develop and validate Cross-Layer Multi-Cloud Application Monitoring- and Benchmarking-as-a-Service Framework (CLAMBS). CLAMBS offer the following novel features:

- It provides the ability to monitor and profile QoS of applications, whose parts or components are distributed across heterogeneous public or private clouds;
- It provides visibility into QoS of individual components on an application stack (e.g., web server, database server). In particular, CLAMBS facilitate efficient collection and sharing of QoS information across cloud layers using a cloud provider agnostic agent-based technique;
- It provides benchmarking-as-a-service that enables the establishment of baseline performance of application deployed across multiple layers using a cloud-provider agnostic technique; and
- It is a comprehensive framework allowing continuous benchmarking and monitoring of multi-cloud, multi-layer hosted applications.

The rest of the paper is organized as follows. Section 2 presents summary of current techniques and frameworks that support cloud monitoring. Section 3 presents the CLAMBS system framework for cloud applications monitoring and benchmarking. Section 4 presents CLAMBS deployment models in multi-cloud environments. Section 5 presents the prototype implementation details. Section 6 presents empirical evaluation results of CLAMBS framework. Finally, section 7 concludes the paper.

## 2 RELATED WORK

In [11], Lattice monitoring framework is presented for monitoring virtual and physical resources. In this paper, a managed service is specified as a collection of Virtual Execution Environment (VEEs). Hence, Lattice is implemented to be able to collect information for CPU usage, memory usage, and network usage of each VEE and VEE host. Moreover, a dependable monitoring facility is presented in [12], called Quality of Service MONitoring as a Service (QoS-MONaaS). The focus of QoS-MONaaS approach is to: (i) continuously monitor the QoS statistics at the Business Process Level (SaaS); and (ii) enable trusted communication between monitoring entities (cloud provider, application administrator, etc.). Furthermore, a monitoring framework known as (PCMONS) is developed by incorporating previous frameworks and techniques [14]. PCMONS proves that cloud computing is viable way of optimizing existing computing resources in data centers. Also, the paper notes that orchestrating monitoring solutions on installed infrastructures is viable. In contrast to above frameworks, CLAMBS focuses on monitoring and benchmarking applications components across cloud layers as well as across heterogeneous cloud platforms. Moreover, current cloud monitoring solutions lacks an integrated approach to benchmarking. For example, cloud harmony[2] makes available a benchmarking performance data of their infrastructure but does not allow end-users to benchmark application components. In [34], authors broadly classify the four areas of benchmarking application in cloud environments as CPU, Memory I/O, Disk I/O, and Network I/O. The proposed CLAMBS framework is driven by these principles of monitoring resources at application component layer cloud layers in multi-cloud environment.

---
[1] http://aws.amazon.com/cloudwatch/

In cloud platforms, recent efforts have been put into improving VMs monitoring and controlling. A number of frameworks have been proposed for VM management, which employ Simple Network Management Protocol (SNMP). SBLOMARS [13] implements several sub-agents called ResourceSubAgents for remote monitoring. Each of SBLOMARS's sub-agents is responsible for monitoring a particular resource. Inside each of these sub-agents, SNMP is implemented for management data retrieval. In contrast to CLAMBS which is focused on monitoring and benchmarking applications QoS in virtualized cloud computing environments, SBLOMAR focuses on enabling multi-constrain resource scheduling in grid computing environments.

In [15], CloudCop is a conceptual network monitoring framework implemented using SNMP. Basically, CloudCop adopts Service Oriented Enterprise (SOE) model. CloudCop framework consists of three components: Backend Network Monitoring Application, Agent with Web Service Clients, and Web Service Oriented Enterprise. While CloudCop focuses on network QoS monitoring, CLAMBS is concerned with application QoS monitoring. In [16], the authors propose a Management Information Base (MIB) called Virtual-Machines-MIB, to define a standard interface for controlling and managing VM lifecycle. It presents SNMP agents, which are developed based on NET-SNMP public domain's agent. Besides read-only objects, Virtual-Machines-MIB provides read-write objects that enable controlling managed instances. To obtain the data of Virtual-Machines-MIB, mostly Libvirt API and other resources such as VMM API are used [16]. While Virtual-Machines-MIB is concerned with monitoring IaaS-level (VM) QoS statistics, it does not cater for the QoS statistics of PaaS level application components.

In [17], the authors stress the importance to have a standardized interface for monitoring VMs on multiple virtualization platforms and this interface should be based on SNMP. The paper presents a framework for VMs monitoring which is fundamentally based on SNMP. The proposed work was built over three different VM hypervisors namely, VMware, Xen, and KVM. These three hypervisor were installed on two different OSs, which are MS Windows and Linux. Similarly to Virtual-Machines-MIB, this framework utilizes Libvirt API. Moreover, it implements an agent extension AgentX using Java. Primarily, this AgentX is to obtain VMs management data for the VMware, Xen, and KVM VMs and eventually the data is presented via web-based management. However, similar to [16], the approach given in [17] focuses on VM-level QoS monitoring, while completely ignoring application component level QoS management and monitoring. In addition to the mentioned works above, libvirit-snmp is a subproject, which primarily provides SNMP functionality for libvirt. Libvirt-snmp allows monitoring virtual domains as well as it allows setting domain's attributes. Furthermore, Libvirt-snmp provides a simple table containing monitored data about domains' names, state, number of CPUs, RAM, RAM limit CPU time.

In cloud environments, traditional benchmarking approaches cannot serve the users' needs [35]. Besides runtime performance, cloud specific attributes such as elasticity, deployment, resiliency, and recovery are required to be reflected in benchmarking process [35]. Further, benchmarking applications distributed in multi-cloud environments is a complex task as each application requires evaluation of distinct QoS metrics from others in order to evaluate the targeted cloud performance. Moreover, each application has its own workload requirements for each individual component rendering the need for a general-purpose benchmark framework. A specific purpose-built benchmarking component will not be able to serve cloud users having variety of use cases in cloud environment.

Authors in [33], put a notable effort on the design and the simplicity of using C-MART, which is, a web application benchmarking tool. C-MART presents a significant tool emulating, and then benchmarking web applications such as online store or social networking website. Originally, C-MART is motivated by the fact that benchmarks need to cope up with the shift from the traditional environments to cloud environments. However, C-MART is limited to benchmarking web application at the PaaS layer.

Amazon EC2 compatible C-Meter was the original prototype of the EC2 current extensible cloud benchmark framework [36]. It employs low level metrics that are typically not visible to general cloud users. Therefore, C-Meter is unsuitable to evaluate higher levels of cloud services (e.g. PaaS and SaaS) [36]. Despite of metrics CloudCmp [37] can measure, authors in [35] stated that some of the metrics provided by CloudCmp are too experimental to be meaningful to cloud user, e.g. time to consistency. CloudGauge [38], presents an effective dynamic virtual machine benchmarking tool. It provides automated scripts to provision and measure the performance of the virtual environment setup. But, the focus of CloudGauge experimental benchmark was on the virtualization layer. Furthermore, the data collected were mainly CPU usage and average load Memory.

To guarantee the SLA and to avoid failure, the challenge is to identify which component of the application needs to be re-configured or what type of auto-scaling is required, To this end, we need a better understanding of individual component's performance accurately to help cloud orchestrator to effectively scale the corresponding layer at the appropriate time. The proposed CLAMBS model benchmarking and real-time monitoring as-a-service system is a practical method to understand and evaluate how application components distributed across cloud layers in multi-cloud environments can essentially perform and handle their tasks.

---

[2] https://cloudharmony.com/services

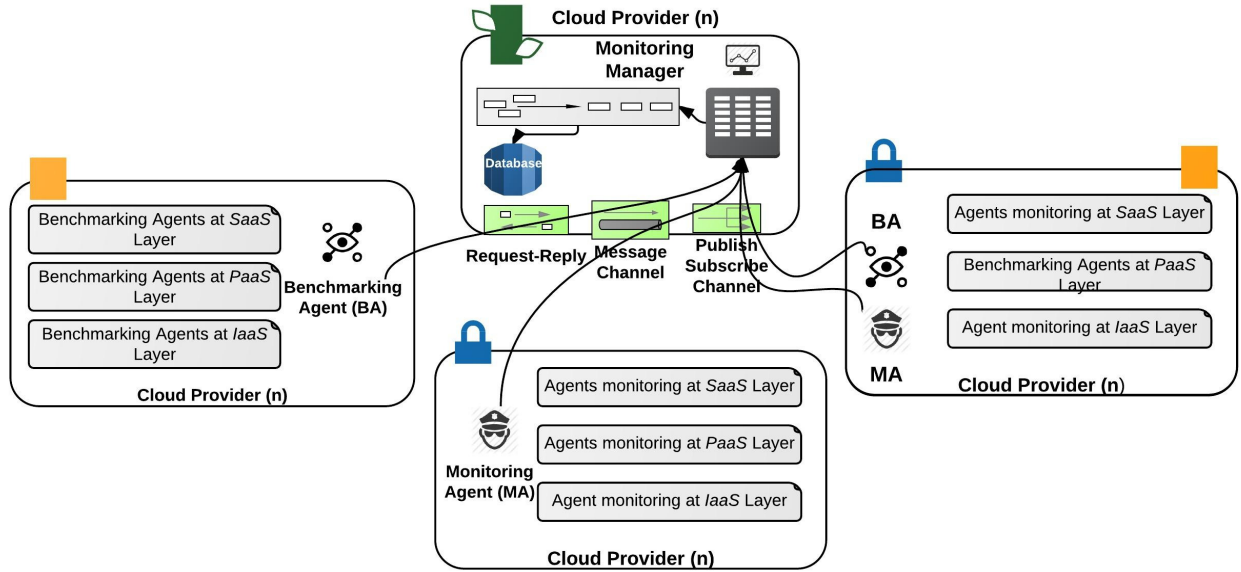

Figure 1: Overview of CLAMBS Model

## 3 CLAMBS: CROSS-LAYER MULTI-CLOUD APPLICATION MONITORING AS A SERVICE

### Overview

Fig. 1 presents an overview of the proposed CLAMBS framework. As depicted in the figure, CLAMBS employs an agent based approach for cross-layer, multi-cloud resource/application monitoring and benchmarking. In this multi-cloud approach, monitoring and benchmarking agents are deployed across various cloud provider environments based on application requirements and deployments.

A CLAMBS agent is responsible for monitoring and benchmarking application QoS parameters such as resource consumption, network performance, storage performance etc at various layers including SaaS, PaaS and IaaS. On the other hand, CLAMBS manager is responsible for orchestrating and collecting QoS data from each monitoring and benchmarking agent.

### CLAMBS Model

CLAMBS include mechanisms for efficient cloud monitoring and benchmarking applications deployed at *aaS layers. CLAMBS provides standard interfaces and communication protocols that enable application/system administrator to gain awareness (benchmark and monitor against benchmarking outcomes) of the whole application stack across different cloud layers in heterogeneous, hybrid environments (different resources constraints and operating systems). The CLAMBS approach also addresses the challenges in interoperability among heterogeneous cloud providers. Fig. 2 presents a detailed architecture of the proposed CLAMBS framework. The CLAMBS framework comprises three main components namely, Manager, Monitoring Agent and Benchmarking Agent.

### A. Manager

The CLAMBS Manager is a software component that performs two operations: 1) it collects QoS information from Monitoring Agents; and 2) it collects benchmarking information from benchmarking agents running on several virtual machines (VMs) across multi-cloud providers and environments. In case of monitoring, the manager collects QoS parameter values from the monitoring agents running at the *aaS layers. The communication between the manager and the agents can employ a push or pull technique. In case of pull technique, the manager polls the CLAMBS monitoring agents at different frequency to collect and store the QoS statistics in a local database (DB).

When a push strategy is employed, the agents obtain the relevant QoS statistics and push the data to the Monitoring manager based on a predetermined frequency. As soon as the monitoring system is initialized in the cloud(s), the VMs running the CLAMBS manager(s) and the monitoring agents boot up. Using discovery mechanisms such as broadcasting, selective broadcasting or decentralized discovery mechanisms [20], the agents and manager discover each other. After discovering the address of each agent and manager, depending on the available strategy (push/pull), QoS statistics is collected by the manager from the agents.

To illustrate further, consider a web multimedia application service hosted on multiple cloud providers for example in US Virginia, and AU Sydney. The users can search the multimedia content and can retrieve the desired content via the web application. Such a web multimedia application comprises the media storage for content distribution at the IaaS layer, a database server for media search and indexing at PaaS and a web interface at the SaaS layer. The media and the database servers are hosted at the PaaS layer, whereas, the media content is stored at the storage server at IaaS layer. Each component of the web application is running and hosted on different VMs. Media server has an IP address say, 192.168.1.1, indexing server has an IP address 192.168.1.2, and the storage server has IP 192.168.1.3. Each VM also runs CLAMBS monitoring agents that monitor applications and VM parameters (e.g. CPU, Storage and Memory). In this case, the manager can send first request to the agent on the media server VM specifying the IP address 192.168.1.1:8000 and stating the QoS target e.g., CPU utilization. Similarly, a second request is sent to the agent on the indexing server VM specifying the IP address (192.168.1.2:8000) and stating the QoS target e.g., Packets In. In the same way, a third request is sent to the agent on the storage server VM specifying the IP address (192.168.1.3:8000) and stating the QoS target e.g. actual used memory.

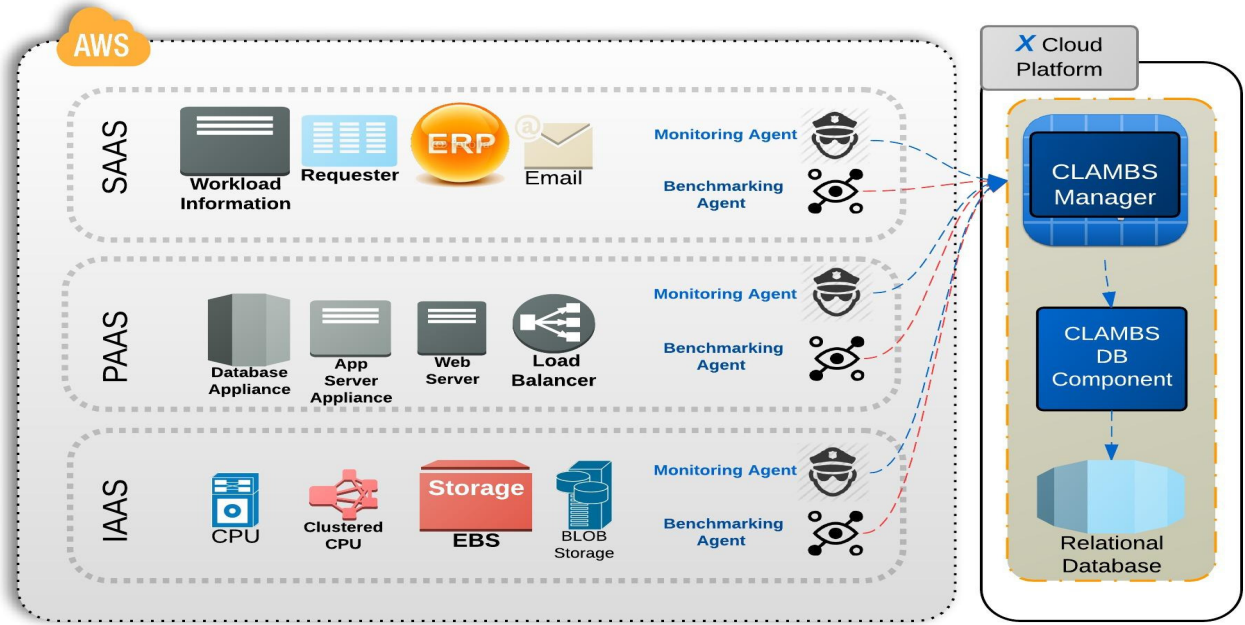

Figure 2: CLAMBS Framework Architecture

The CLAMBS manager employs a QoS data collection schema to store QoS statistics collected from monitoring agents into the local database and an agent schema to maintain the list of discovered agents. The second operation is performed by the CLAMBS Manager to facilitate benchmarking of applications distributed across *aaS layers in multi-cloud environments. The manager's benchmarking function is a software component that collects network and application performance QoS information from CLAMBS benchmarking agents that are distributed and running on several VMs hosted across multi-cloud environments in different data centers. In particular, the manager collects the traffic QoS values from agents hosted on VMs that are distributed across different data centers.

The benchmarking component of the manager is responsible for firing VMs at remote data centers to perform application level benchmarking based on user requirements that include data center locations. For example, consider a scenario where an end user located in Singapore, requests multimedia content from the web multimedia application service. Typically, such application components could be distributed across multiple datacenters. The CLAMBS framework supported by the manager is able to dynamically fire a VM hosting the benchmarking agent at the end user location, Singapore. Then the CLAMBS manager can repeatedly test and benchmark the performance of the web application at both the locations (US Virginia, and AU Sydney) to select the best location to serve the multimedia content to the end user. This approach serves the following two main purposes: 1) it allow users who use third-party cloud hosting services to benchmark application performance for later comparison and evaluation; and 2) it allow users to test the system's performance automatically and choose the best performing data center for service delivery. The key advantage of CLAMBS here is the ability to dynamically run benchmarking of application at *aaS layers of multiple clouds automatically with very little configuration required from the user. The CLAMBS manager also incorporates an API that is used by other monitoring manager or external service to share the QoS statistics.

### B. CLAMBS Monitoring Agent

Another major component of the CLAMBS framework is the monitoring agent. The monitoring agent resides in the VM running the application and collects and sends QoS values as requested by the manager. After the monitoring system initialization, the agent waits for the incoming requests from the manager or starts to push QoS data to the manager. Upon arrival of the request, the agent retrieves the stated QoS values belonging to a given application process and/or a system resource and sends them back as a response to the manager.

The monitoring agent has the capability to work in multi-cloud heterogeneous environments. Agent manager communication can be established using any approach that fits the application requirement e.g., publish- sub-scribe, client- server or web services. It can also employ standardized protocols for communicating system management information like SNMP. The proposed blueprint does not restrict future developers from extending CLAMBS to their purposes. In our proof-of-concept implementation explained later, we demonstrate the implementation of the CLAMBS framework using a combination of SNMP and RESTful Web services. The CLAMBS monitoring agent also uses operating system dependent code to fetch corresponding application QoS statistics, for example, use of OS specific commands to get CPU usage in Linux and Windows systems respectively.

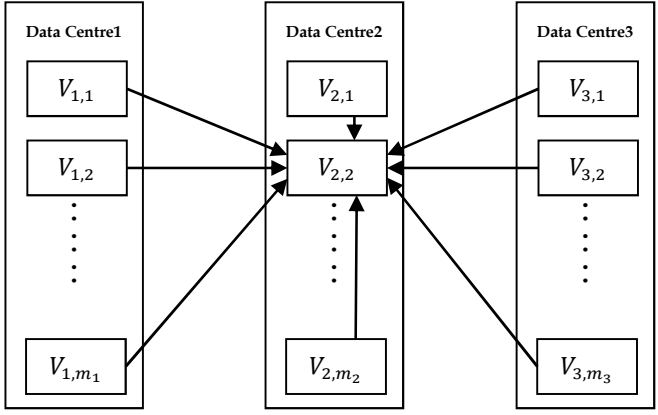

Figure 3.1: Communications: 3 data centers, manager $\theta$ located on $V_{2,2}$

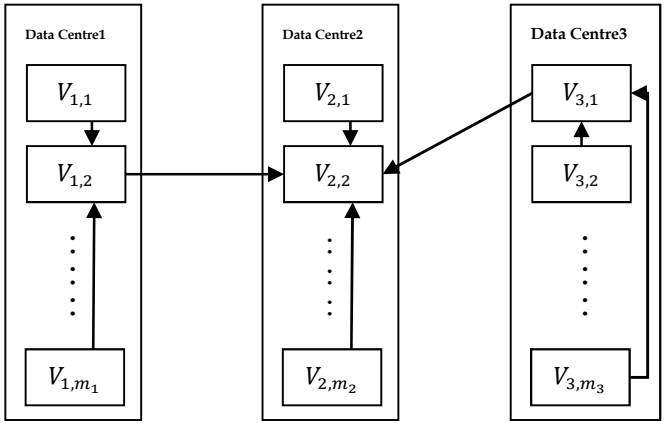

Figure 3.2: Communications: 3 data centre, manager $\theta$ located on $V_{2,2}$

### C. CLAMBS Benchmarking Agent

The third component of the CLAMBS framework is the benchmarking agent. This agent has the capability to migrate from the manager VM to a VM that either hosts the application/service or act as a client to the service. The benchmarking agent incorporates standard functions to measure the network performance between the data center(s) hosting the application service and the client. The benchmarking agent also incorporates a load-generating component that generates traffic to benchmark the application based on a workload model. The load generator part of the benchmarking agent is able to generate load on applications such as DBMS and Web Servers. For example, generating requests to a web server (*N* users and *M* requests/second) based on a website workload model (e.g. football world cup trace - http://ita.ee.lbl.gov/html/contrib/WorldCup.html). The benchmarking agent has the capability to work in multi-cloud heterogeneous environments.

In essence, objectives that require benchmarking process are: i) determining where and what type of performance improvements are needed, ii) analyzing the available metrics of performance, iii) using benchmarking information order to improve the services performance, and iv) comparing the benchmarking information with the standard measurements. Thus, to benchmark cloud applications (e.g. web application), providers can apply a workload on such application's distributed components. Compared to the state-of-the-art research, CLAMBS benchmarking functionality is an additional dimension alongside monitoring. This means that CLAMBS is one of a kind unified framework incorporating monitoring and benchmarking as-a-service capabilities based on distributed agents across multi-cloud platforms.

## 4 MODELING AND ANALYZING CLAMBS OVERHEADS IN MULTI-CLOUD ENVIRONMENTS

As mentioned previously, the CLAMBS monitoring framework is aimed to be agnostic of the underlying cloud platform i.e., the manager/agent may run on heterogeneous cloud platforms. In case the monitored framework is distributed across different cloud platforms e.g., Amazon cloud platform and Windows Azure platform, then one manager and multiple agents will be residing on each of these cloud platforms. Hence, it is important to model the overheads introduced by the distribution of CLAMBS in multi cloud environments.

**Communication Overhead**

The communication overhead depends on the physical locations of managers i.e., data center where CLAMBS Agents are distributed across different data centers. We have $n$ data centers $D_1, D_2, \ldots, D_n$. For a data center $D_i$, there are $m_i$ VMs running: $V_{i,1}, \ldots, V_{i,m_i}$. As each VM is accompanied by a CLAMBS agent, we denote the agents as $\forall_i = \forall_{i,1}, \ldots, \forall_{i,m_i}$. The size of one CLAMBS Agent message from $\forall_{i,j}$ is $M_{i,j}$. Location and deployment of CLAMBS agents and managers will vary. When there is one CLAMBS manager $\theta$ located on data center $D_\pi, \pi \in [1, n]$ (See Fig.3.1): each of the agent $\forall_{i,j}$ VM has to communicate with the manager independently; thus, total communication overhead from CLAMBS agents to CLAMBS manager in one report will be as following:

$$\sum_j \left( \left( \sum_i M_{i,j} \right) - M_{\pi,j} \right), i = 1, \ldots, n; j = 1, \ldots, m_i \quad (1)$$

If the message size is a fixed value M then CLAMBS messages communication overhead is

$$M \cdot \left( \left( \sum_i m_i \right) - m_\pi \right), i = 1, \ldots, n \quad (2)$$

In the above formulas, messages of agents located in π are excluded being in the same data center where CLAMBS Manager is running. Furthermore, for optimiza-tion, these messages may not be needed for every report. This will take place when CLAMBS agent process data analysis before sending data. Therefore, when changes occur to data then they will be reported to CLAMBS manager. Thus, If only a subset $S_i$ of $\forall_i$ is reporting each time, CLAMBS communication cost will be reduced greatly.

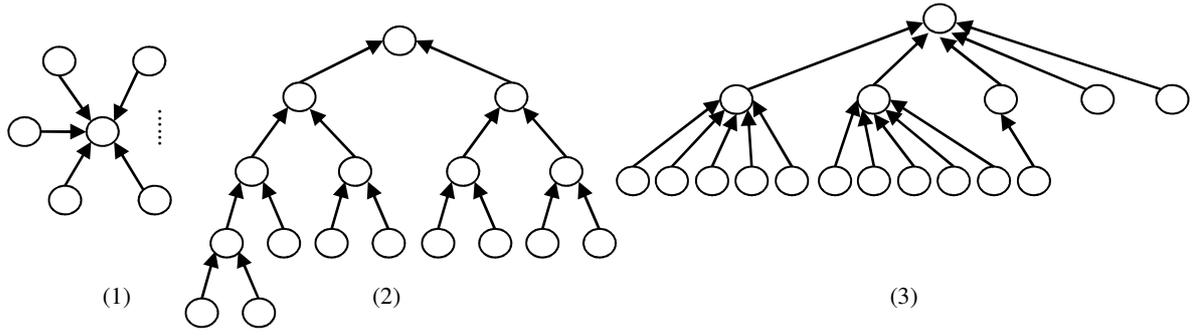

Figure 4. Different management structures for 17 agents

Let $\Pi_i$ be the bandwidth (connection speed) for data center $D_i$. The total time consumption in communication (when all CLAMBS messages are sent simultaneously at fixed time slots) is:

$$\text{MAX}_i\left(\text{MAX}_j\left(\Pi_i \cdot M_{i,j}\right)\right) \quad (3)$$

$$\text{MAX}_i\left(\text{MAX}_j\left(M_{i,j}/\Pi_i\right)\right) \quad (4)$$

Therefore it is possible to develop adaptive algorithms to reduce reports from agents $\forall_{i,j}$ with large $\Pi_i \cdot M_{i,j}$ to save time, at the cost of CLAMBS messages info. As they are all variable, the criteria could be an average from history. This is a possible way to decide $S_i$ for every agent report.

When there are $n$ distributed CLAMBS managers/sub-managers located across different data centers (See Fig. 3.2), the cost is significantly reduced. Ideally, $n$ managers $\theta_1, \theta_2, \dots, \theta_n$ are located in different data centers. Although management task is distributed, a super manager is still needed for maintaining a centralized database. Let's say the super manager is $\theta_\pi \in \{\theta_1, \theta_2, \dots, \theta_n\}$. In this case, if the message size from $\theta_i$ is $M_i$, then the total communication overhead for each round is reduced to $\sum_i M_i$. However, the optimization in communication overhead also brings other trade-offs or compromises such as in setting up and switching additional managers, CPU load, response time, etc. We now discuss further in the following section.

**CPU, Response and Search Time**

The distributed CPU load will be determined by the layout of agents. We will also compare the standard one-manager layout (model (1), see Fig.4.1) against the hierarchical tree-typed manage structure (model (2 & 3), see Fig.4.2, 4.3). The total number of agents is $N$ and the max number of child nodes per node is $n$. The CPU load for managing one CLAMBS message is $C$. If there are a total of $l$ levels of the tree control structure, then:

$$l \geq \lceil \log_n(N \cdot (n-1) + 1) \rceil \quad (5)$$

the inequality turns into an equality when the tree is a complete tree in its top $l-1$ levels. In model (1) (Fig 6.1), CPU load for the super manager per round is $(N-1) \cdot C$ and other nodes is 0. In model (2) (Fig 6.2), max CPU load for super manager will be $\cdot C$, and at least $\lceil (N-1)/n \rceil$ other managers will also take over a maximum CPU load of $n \cdot C$ each. Whatever the load distribution, as the same total number of agents are returning the same amount of CLAMBS data, the overall CPU load will remain the same. In other words, a larger $n$ will incur less managers to participate and increase the load for each manager. Smaller $n$ will improve the distribution, but $l$ will also increase so that the response time will grow.

The response time will be determined by the time for a node used to reach super manager for it to react on unusual behaviors. If the time for a node (agent) $\forall$ to contact its manager is $t$ (including processing and communication), then in (1) all response time is $t$. In (2), the response time will grow for most nodes. The response time for node $\forall$ will be $l_\forall \cdot t$ where $l_\forall$ is the level of $\forall$. Under this model, it's easy to observe that a larger $n$ will cause less number of higher-response-time nodes, therefore smaller total response time. As the response time for most individual nodes will grow, the total response time for all $N$ nodes will also grow. Instead of $(N-1)t$, the total time $t_{\text{total}}$ satisfies:

$$t_{\text{total}} \geq t \cdot \left(\sum_{i=1}^{l-2} i \cdot n^i + (l-1) \cdot \left(N - \sum_{j=0}^{l-2} n^j\right)\right)$$

$$= t \cdot \left(\frac{(l-2)n^l - (l-1)n^{l-1} + n}{(n-1)^2} + (l-1) \cdot \left(N - \frac{n^{l-1} - 1}{n-1}\right)\right)$$

$$= t \cdot \left((l-1) \cdot N - \frac{n^l - ln + l - 1}{(n-1)^2}\right) \quad (6)$$

Therefore, the average response time $t_{\text{avg}}$ for $N-1$ nodes other than the super manager satisfies

$$t_{\text{avg}} \geq \frac{t}{N-1} \cdot \left((l-1) \cdot N - \frac{n^l - ln + l - 1}{(n-1)^2}\right) \quad (7)$$

As before, the inequalities turn into equalities if and only if the tree is a complete tree in the top $l-1$ levels. We can see that given a fixed $N$, when $n$ decreases or $l$ increases, the average response time will grow. Note that here $t$ is considered a constant value. In practice, communication overhead will also affect response time of each node. Therefore, minimizing inter-data center communications as shown in communication overhead analysis will also help in a lowering response time.

Another metric is the average search time. Similar to a search tree, the (minimum) average search time for the super manager to find a leaf node in (2) is $\log_n N$ (for a complete tree), as opposed to $((N-1)/2) \cdot t$ in (1). Therefore, the search time will also benefit from a larger $n$.

To sum up, we can see that the two deployments of agents have their own advantages and disadvantages. To achieve deserved performance, the system setup will depend on the actual requests and different metrics such as communication overhead, CPU load distribution, average response time analyzed in this section.

## 5 CLAMBS: SYSTEM IMPLEMENTATION

The proof-of-concept implementation of the proposed CLAMBS framework has been developed using Java and is completely cross-platform interoperable i.e., it works on both Windows and/or Linux operating systems.

*Monitoring Agent Implementation*: The process of retrieving QoS targets is done by utilizing functionalities provided by SNMP, SIGAR, HTTP and other custom built APIs. For instance, SNMP is used to retrieve the QoS values related to networking, number of packets in and out, route information and number of network interfaces. SIGAR is used to obtain access to low-level system information such as CPU usage, actual used memory, actual free memory, total memory and process specific information (e.g. CPU and memory consumed by a process). Moreover, network information such as routing tables can also be obtained using SIGAR. Both SIGAR and SNMP packages have their own operating system specific implementations to retrieve system information e.g. system resources, and user processes. To enable SNMP monitoring, we define new SNMP Objects Identifiers (OIDs) in a sequence. For example function to get the CPU usage of a specific process (tomcat) is assigned an OID .1.3.6.1.9.1.1.0.0. Similarly, function to get process memory is assigned an OID .1.3.6.1.9.1.1.0.1. The CLAMBS implementation also incorporates a HTTP based Restlet communication standard. This allows greater flexibility to monitor application that does not support the network specific SNMP protocol.

*Manager Implementation*: The manager uses a MySQL database to store the QoS statistics collected from the monitoring and benchmarking agents. For the proof-of-concept implementation, we used a pull approach where the Manager is responsible to poll for QoS data from agents distributed across multiple cloud provider VMs. The manager uses a simple broadcasting mechanism for agent discovery. On booting, a discovery message is broadcasted to the known networks. Agents that are available respond to the manager's request. The manager then records agent information to the agent database. The manager then starts off threads to query each agent in the agent database to obtain QoS parameters. The polling interval is a pre-defined constant and can be changed using the manager configuration files. Utilizing Java functionalities, the manager is implemented based on the net package which is provided by Java libraries. This library is responsible of most network communication functions and requirements. It provides the superclass URLConnection which represents a communication link between applications and Uniform Resource Locator (URL). Therefore, each manager's request will have two main components which are protocol identifier and resource name. The benchmarking component of the manager can measure the QoS parameters including Network Latency, Network Bandwidth, Network download speed, and Network upload speed. We have also incorporated RESTful-based API's allowing external services/applications to query monitoring and benchmarking data.

*Benchmarking Agent Implementation*: Benchmarking agents are bootstrapped with the VMs and distributed across different cloud platforms e.g. Amazon and Azure. On booting VMs, agents start up and wait for incoming requests from the manager to start benchmarking. Typically, there is a unique IP address for each agent representing the VM location. The port used for communication by the benchmarking agents is 80 as the protocol identifier in our implementation for communication is HTTP. The server component we integrate to run the agents is Apache Tomcat. Upon requests by the manager, the agent starts its role which includes download/upload objects from remote server. Essentially, the agent is capable of handling requests from more than one sub-manger in case of hierarchal architecture are adopted where sub-manager and one super manager are in use. The benchmarking agent also incorporates the load generator. This component of CLAMBS is essentially implemented using the JMeter package developed in Java. In this implementation we designed our prototype to generate web application server traffic using HTTP requests. The system also supports SQL load generation. In case of HTTP workload, HTTP sampler is provided along with the domain, port number, path, and the request method (e.g. POST or GET). Similarly, in case of the SQL workload, SQL sampler, query, query type (insert, update, or select), database URL, and database driver are provided. Loop controller is specified according to the aimed workload scenario. This also applies to the thread group and the number of threads that will perform the intended workload. Seamlessly, CLAMBS load generator prototype is implemented to be able to reach the targeted components across different cloud platforms.

*Agent Manager Communication*: For the proof-of-concept implementation, the communication between the agent and the manager has been implemented using two techniques namely RESTful Web services and SNMP. Having a RESTful approach enables easy lightweight communication between CLAMBS agents and manager/super manager. Using a standardized SNMP interface makes CLAMBS completely compatible with existing SNMP-based applications, tools and systems and reduces the effort involved in collecting QoS statistics.

## 6 CLAMBS: EXPERIMENTS AND RESULTS

### Hardware and Software Configuration

To evaluate the CLAMBS framework, experiments were conducted on Amazon AWS and Microsoft Azure platforms. We used standard small instances on each platform. The AWS instance has the following configurations: 619 MB main memory, 1 EC compute unit e.g., 1 virtual core with 1 EC2 compute unit, 160 GB of local instance storage, and a 64-bit platform. The Azure instance has the following configurations: 768 MB main memory, 1GHz CPU (Shared virtual core) and a 64 bit platform. Three different data centers are considered in this experiment, namely, Sydney, US-Virginia, and Singapore. CLAMBS Manager was located in Sydney. One CLAMBS agent was hosted on a VM at US-Virginia data center and another CLAMBS agent is hosted on a VM in Singapore data center. VM's in the experiments were running Microsoft Windows Operating System. For persistent storage of CLAMBS agent and manager data, we used off storage volumes such as Elastic Block Store (EBS) in Amazon EC2 and XDrive in Windows Azure. Major advantages of architecting applications to adopt off instance storage are: i) each storage volume is automatically replicated, and this prevents data loss in case of failure of any single hardware component; and ii) storage volumes offer the ability to create point in time snapshots, which could be persisted to the cloud specific data repositories.

**Experimental Setup**

As discussed previously, the CLAMBS system has three main components namely the Manager, Monitoring agent and Benchmarking agent. In this section, we present the experimental scenario and setup of the monitoring and benchmarking agents. In both cases, the manager is responsible to collect monitored and benchmarked QoS parameters.

To evaluate and validate CLAMBS system, we consider a web multimedia application that uses a content distribution network to distribute multimedia content to end-users using a multi-cloud provider setup (e.g. combination of Amazon AWS and Windows Azure). We employ CLAMBS approach to benchmark and monitor the performance of the web multimedia application components namely the search and indexing server (Tomcat web server and MySQL database) and network QoS parameters including network latency and download and upload performance.

*Monitoring Agent Setup:* Each monitoring agent comprises the corresponding SNMP and SIGAR package dependencies to accomplish the monitoring task. In the experiment, the monitoring manager triggered a request to monitoring agents, which in turn retrieved the requested QoS parameters from the hosted VM. Each agents running on the VM listened on a unique port e.g. VM1-IP:8000, VM1-IP:8001, enabling them to respond to queries from the monitoring manager independently. The agents sent responses to the monitoring manager concurrently. For experimental purposes and to demonstrate and validate CLAMBS cross-layer monitoring capability, each agent monitored several resources including system resources and user processes

Table 1 presents the list of monitored processes/resources. On retrieving QoS data from the agents, the monitoring manager saved the data into the local database by classifying them as system performance or user applications QoS performance parameters.

Table 1: Monitoring across different layers

| Process/Resource | Description | Owner |
|---|---|---|
| Tomcat7w.exe | Apache Tomcat 7 | User |
| MySqld.exe | MySQL Workbench 6.0 | User |
| Javaw.exe | Monitoring Manager | User |
| Lsass.exe | Local Security Authority Process | System |
| Winlogon.exe | Windows Logon App. | System |
| Services.exe | Services and Controller App. | System |
| VM CPU Usage | CPU usage of the entire VM | System |
| VM Memory Usage | Memory usage of the entire VM | System |

*Benchmarking Agent Setup:* The benchmarking agent is composed of two components which are network traffic benchmarking and CLAMBS load generator. Each agent comprises the corresponding required Java packages dependencies to accomplish the benchmarking task. In this experiment setup, we test the network QoS parameters that links the CLAMBS manager and the benchmarking agents. Benchmarking the network link connecting an agent and the CLAMBS manager was accomplished by generating bi-directional traffic to simulate download and upload processes. We ran this experiment to demonstrate CLAMBS ability to benchmark network performance between two different locations of data centers.

In our experiments, the CLAMBS manager triggered the benchmarking requests to CLAMBS benchmarking agents, which responded immediately to the manager's request. Communications between CLAMBS manager and agents were conducted using the RESTful HTTP protocol. Pre-defined files with varying sizes (50 MB, 100MB, and 200MB) were used during the experiment to measure network performance over a download/upload processes. Table 2 lists the measurements parameters that were observed throughout the experiment. According to our conceptual framework, such measurements provide the user with the ability to decide and choose a preference of what site/location a service is performing better. Likewise, a service provider will certainly acquire such knowledge in order to improve the delivered service quality to clients.

*Runtime Configuration Monitoring Agent*: Monitoring agents as well as manager are packaged into jar files with corresponding dependencies and configured to run during VM boot process. The agents use a configuration file that specifies processes to monitor. Based on this information, at run-time, the agent determines the process id of the respective process. After finding the process id, the agent starts to retrieve specific QoS parameters for that process e.g. memory usage and CPU consumption.

Table 2: Benchmarking Measurements

| Traffic Benchmarking Measurement Parameter | Description |
|---|---|
| Download File Network Latency Time | Time consumed starting from a request up-till download complete including Network Latency |
| Upload Network Bandwidth | Amount of data transferred per Second while download process |
| Upload File Network Latency Time | Time consumed starting from a request up-till upload complete including Network Latency |
| Upload Network Bandwidth | Amount of data transferred per Second while upload process |

Fig. 5 provides a detailed workflow of communication between the monitoring manager and agent. The monitoring manager instantiated parallel threads for each group of Agents in one VM i.e., each thread was dedicated to only one VM to communicate with Agents running on that VM. Manager thread sent requests to agents addressed by IP address and port number. The request was for a list of QoS parameters monitored by the agent. After receiving the request, agents compute the QoS parameter values from the hosting VM. The agents then respond to the manager with corresponding QoS parameters.

To evaluate the proposed CLAMBS framework, we deployed the agents and managers on four virtual machine instances (3 VM's on AWS and 1 on Microsoft Azure). On VM's that hosted the agent, depending on number of agents, the agents were bound to unique ports. E.g., if VM-3 hosted 30 Agents, it was bound to ports 8000-8030. Similarly if VM-4 hosted 10 agents, it was bound to ports 8000-8010.

*Runtime Configuration Benchmarking Agent*: CLAMBS manager and agents are packaged into runnable jar and war files with corresponding dependencies and configured to run during VM boot process. The agents use a configuration file that is required to run and remain standby waiting for the manager requests. Intervals of requests can vary but initially is set to 10 seconds for each request sent to a single CLAMBS agent. Agents in turn take immediate response towards CLAMBS manager request. Definite data with pre-chosen sizes are stored locally in each VM hosting CLAMBS manager and CLAMBS agents to be utilized for data transferred during the experiment. Fig. 6 provides a detailed workflow of communication between the CLAMBS manager and agents in different data centers. The manager instantiated parallel threads for each agent addressed by IP address and the port number . Concurrently, CLAMBS manager send similar requests to other registered agents in different data centers which can also be in a different cloud platform.

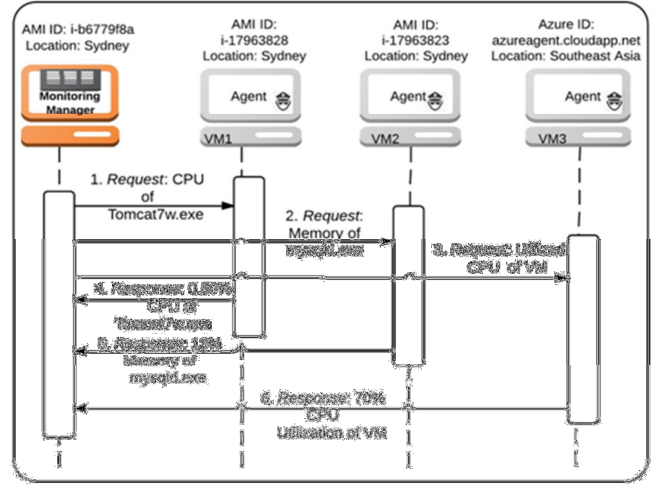

Figure 5: Manager/Agents run-time workflow

## Experimental Results and Discussion

### CLAMBS Monitoring Agent

To validate that the CLAMBS monitoring agent does not introduce significant overheads while monitoring QoS parameters across layers in multi-cloud environments, we ran experiments in 4 typical multi-cloud workload scenarios.

**Scenario I**: VM-1 hosts the Manager, VM-2 hosts 25 Agents, VM-3 hosts 30 Agents, and VM-4 hosts 30 Agents. In total, the manager communicates with 85 Agents deployed in multi-cloud environment (3 AWS instances and 1 Azure instance).

**Scenario II**: VM-1 hosts the manager, VM-2 hosts 10 agents, VM-3 hosts 20 agents, and VM-4 hosts 20 agents. In total, the manager communicates with 50 Agents.

**Scenario III**: VM-1 hosts the manager, VM-2 hosts 10 Agents, VM-3 hosts 10 Agents, and VM-4 hosts 10 Agents. In total the manager communicates with 30 Agents.

**Scenario IV**: VM-1 hosts the manager, VM-2 hosts 1 agent, VM-3 hosts 1 agent, and VM-4 hosts 3 agents. In total the manager communicates with 5 Agents.

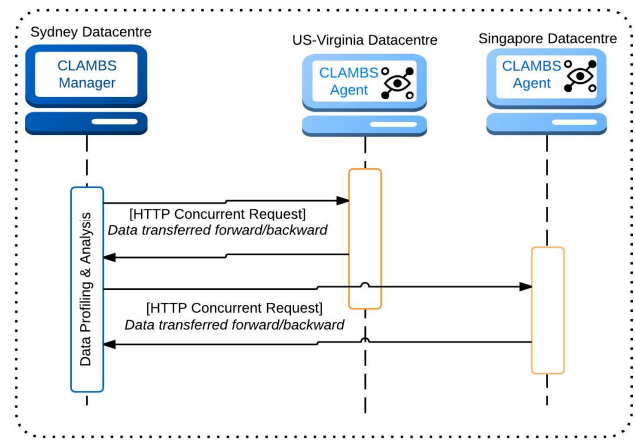

Figure 6. CLAMBS Benchmarking components communication

For each scenario, we monitored the CPU and memory consumption of the monitoring manager. The result of the experiments is presented in Fig. 7 and 8. We computed the average CPU and memory utilization by the Manager for each scenario. Each evaluation scenario involving communication between agents and manager was run for duration of 30 minutes. The frequency of querying the agents for QoS parameters was set to 1 second. The outcomes clearly indicate that the manager performance is stable with increase in the number of active agents. The CPU utilization grows up from 6.25% when manager is communicating with 5 Agents to 10.92% when the number of agents is 85. Likewise, memory consumed by the manager increased marginally from 177.5 MB with 5 agents to 177.85 MB with 85 agents. Moreover, we note, the manager or the agents during the experiment did not encounter any crash or malfunction. These outcomes clearly validate the resource efficient operation of the CLAMBS framework and its ability and suitability to scale across multi-cloud environments.

In essence, we are motivated by the fact that there is a need for monitoring specific processes across cloud layers in multi-cloud environments. The proposed framework namely CLAMBS demonstrates its capability to achieve this goal by enabling cross-layer monitoring in multi-cloud environments. Experimental evaluations of the CLAMBS framework show a steady scalability of the monitoring manger while handling data from 5, 30, 50 and 85 agents simultaneously. Additionally, we note that the resource requirements of the CLAMBS agent did not increase significantly when testing in environments with 5 and 85 agents.

### CLAMBS Benchmarking Agent

To demonstrate CLAMBS benchmarking ability, we benchmark the network performance between data centers in different locations based on the experimental setup presented earlier.

*Data Download Latency-* Concurrently, CLAMBS manager starts downloading data from agents in Singapore and US-Virginia data centers. Each request indicates what size of data is to be downloaded (50MB, 100MB, or 200MB). As presented in Fig. 9, CLAMBS agent in Singapore data center provided faster data download comparing to CLAMBS agent in US-Virginia. Moreover, we observed that as the data size increase, the data transfer latency from CLAMBS agent in US-Virginia also increases. Such observations are expected to have a major impact on both service provider and service client.

*Data Upload Latency* – experiments as shown in Fig. 10 demonstrates how network traffic benchmarking has the potential to drive preferences of both service provider and service client. Uploading 50MB, 100MB, and 200MB files from Sydney to Singapore show shorter latency times comparing to uploading the same size of data to US-Singapore.

*Download/Upload Bandwidth* – experimental results as shown in Fig. 11, presents the outcome of upload/download bandwidth between Singapore, Sydney and US-Virginia. With 50MB, 100MB, and 200MB size of data being transferred, network bandwidth between Sydney and Singapore remains the same at 8 KB/s. Similarly, the network bandwidth between Sydney and US-Virginia is 6 KB/s for the different data sizes transferred. Although, this is basically a proof-of-concept where the CLAMBS benchmarking capability enables the user to prefer a location over another, in our experimentation scenario Singapore site measured significantly better performance over US-Virginia.

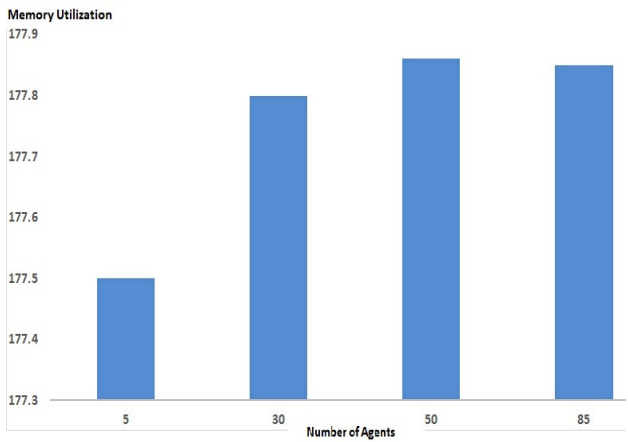

Figure 7: Manager Memory Utilization in MB

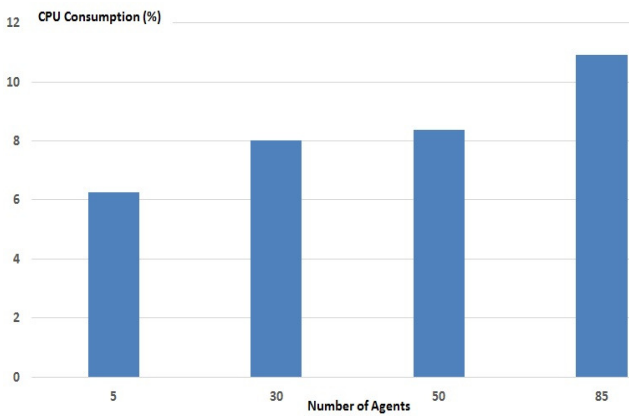

Figure 8: Manager CPU Utilization in Percentage

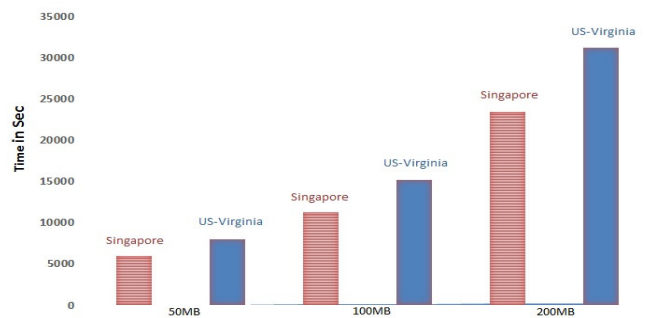

Figure 9. Data Download Network Latency (Time in Seconds)

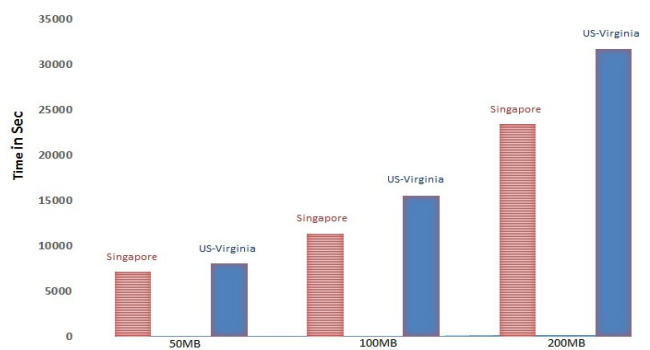

Figure 10. : Data Upload Network Latency (Time in Seconds)

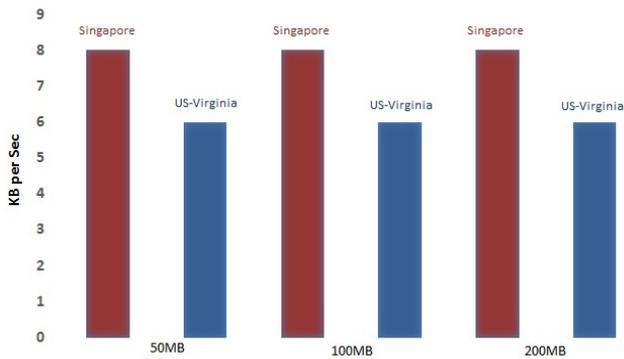

Figure 11: Download/Upload Bandwidth (Kilobytes per Seconds)

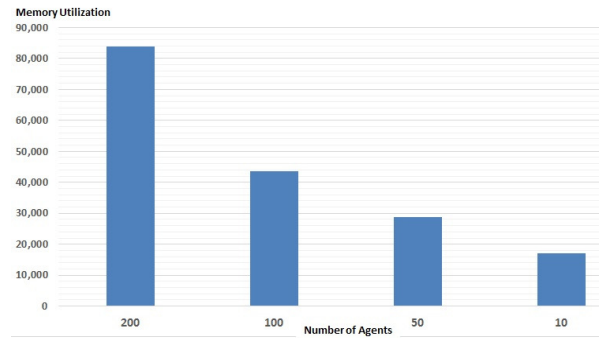

Figure 12. CLAMBS Manager memory consumption (benchmarking scenario)

*Analysis* - Referring to AWS documentation, network performance for small instance types are low. Moreover, such types of instances are not listed under eligible instances for enhanced network performance. Unlike other instance types (e.g. c3.large, c3.xlarge, c3.2xlarge, c3.4xlarge, c3.8xlarge, i2.xlarge, i2.2xlarge, i2.4xlarge, i2.8xlarge, r3.large, r3.xlarge, r3.2xlarge, r3.4xlarge, or r3.8xlarge), small instance type does not have a feature of enabling enhanced network performance. This limitation was reflected by our experiments by having low network bandwidth across different data centers. Furthermore, VM requests serving priority by the hosting server at Amazon platform is low which means that the performance is minimal for such small instances.

*CLAMBS Manager Scalability under Benchmarking-* We also computed the average CPU and memory utilization by the CLAMBS Manager while performing benchmarking of application's network performance. We used a file size of 100 KB enabling us to repeat the operation of data transfer between manager and agents located in different in remote data centers locations. In this scenario, we utilized the CLAMBS monitoring agents to monitor the performance of the CLAMBS manager. Fig. 12 shows the outcome of our experiments. As indicated by the experimental outcome and similar to the Manager's performance while monitoring, the overheads imposed by the benchmarking component of the manager on the underlying system memory consumption is not very significant. The CPU consumption of the manager during benchmarking scenario was also not significant and ranged between 2 – 5%.

The experimental outcomes validate the CLAMBS framework's ability to be a reliable, resource efficient cross-layer monitoring and benchmarking system that can scale across multiple cloud provider environments.

## 7 CONCLUSION

This paper presented CLAMBS, a novel cross-layer multi-cloud application monitoring and benchmarking as-a-service framework. CLAMBS enables efficient QoS monitoring and benchmarking of cloud application components hosted on multiple clouds and across multiple cloud layers. Using experimentation and prototype implementation, we show that CLAMBS is flexible and resource efficient and can be used to monitor several applications and cloud resources distributed across multiple clouds.

As future work, we intend to integrate CLAMBS within a cloud orchestration framework to provide QoS-awareness for cloud admission control and scheduling of Big Data applications in a highly distributed multi-cloud environment.